\documentclass[fleqn,10pt]{wlscirep}
\usepackage[utf8]{inputenc}
\usepackage[T1]{fontenc}
\usepackage{multirow}

\title{Federated Learning for Multi-Center Imaging Diagnostics: A Study in Cardiovascular Disease}
\author[1,*]{Akis Linardos}
\author[1]{Kaisar Kushibar}
\author[2]{Sean Walsh}
\author[1]{Polyxeni Gkontra}
\author[1]{Karim Lekadir}
\affil[1]{University of Barcelona, Department of Mathematics and Computer Science, Barcelona, 08007, Spain}
\affil[2]{Radiomics, Liege, 4000, Belgium}

\affil[*]{linardos.akis@ub.edu}

\begin{abstract}
Deep learning models 
can enable accurate and efficient disease diagnosis, but have thus far been hampered by the data scarcity present in the medical world. Automated diagnosis studies have been constrained by underpowered single-center datasets, and although some results have shown promise, their generalizability to other institutions remains questionable as the data heterogeneity between institutions
is not taken into account.
By allowing models to be trained in a distributed manner that preserves patients' privacy, federated learning promises to alleviate these issues, by enabling diligent multi-center studies.  
We present the first federated learning study on the modality of cardiovascular magnetic resonance (CMR) and use four centers derived from subsets of the M\&M and ACDC datasets, focusing on the diagnosis of hypertrophic cardiomyopathy (HCM).
We adapt a 3D-CNN network pretrained on action recognition and explore two different ways of incorporating shape prior information to the model, and four different data augmentation set-ups, systematically analyzing their impact on the different collaborative learning choices. 
We show that despite the small size of data (180 subjects derived from four centers), the privacy preserving federated learning achieves promising results that are competitive with traditional centralized learning. We further find that federatively trained models exhibit increased robustness and are more sensitive to domain shift effects.

\end{abstract}
\begin{document}

\flushbottom
\maketitle
%
%
\thispagestyle{empty}


\section{Introduction}

Diagnostic tools based on artificial intelligence models have shown promising results in a variety of single-center studies across multiple medical imaging domains~\cite{zhou2021review}, but their generalizability to unseen distributions remains understudied, and their application in clinical practice is still far from realized. As data remains segregated in different institutions, such studies have mostly focused on limited single-center datasets for their training and evaluation. Aside from the obvious issue of having a small sample size, evaluation in such a set-up is questionable, as no assumptions can be made on how this performance translates to unseen centers. For ML methods to generalize to unseen datasets, it is often assumed that newly seen data is independent and identically distributed (IID) to the one seen during training---i.e. each data point comes from the same probability distribution and is mutually independent to all others. For this reason, the data heterogeneity present in multi-center medical data poses a significant problem, as in all cases such data is non-IID---a direct result of the usage of different acquisition protocols, different scanners and varying demographics. 

\begin{figure}[t]
\begin{center}
\includegraphics[width=\textwidth]{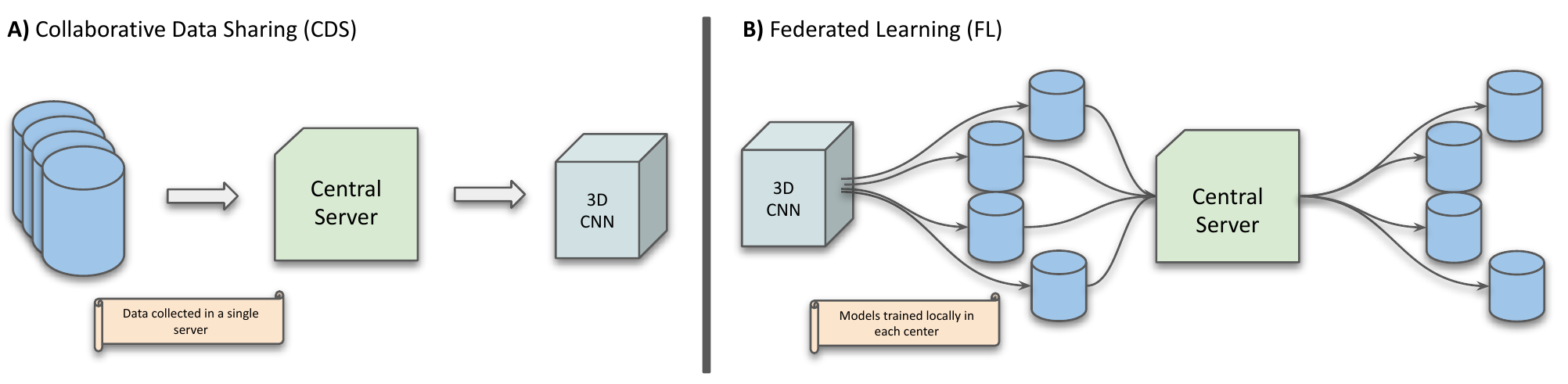}

\caption{A diagram of the two collaborative learning frameworks. For FL only models are transferred, while in the CDS case, the data itself is transferred to the central server and all training happens there.}
\label{fig:cdsvsfl_graph}
\end{center}
\end{figure}

To overcome the core obstacle of data scarcity and to better understand the effects of data heterogeneity that is present in between different centers, institutions need to come together in collaboration. This has thus far been difficult, as institutions are inclined to keep a tight grip on their medical data due to privacy regulations (e.g. GDPR in the European Union and HIPAA in the United States). While an obvious approach for collaborators would be to share their data on a central server (CDS, Figure~\ref{fig:cdsvsfl_graph}A), this endangers patients' privacy by increasing the chance of data leakage. Distributed learning allows for AI models to be trained across multiple edge devices or centers, without data ever leaving its original place~\cite{boyd2011distributed, lambin2015modern}. In 2017, Google proposed Federated learning~\cite{FEDERATEDORIGINmcmahan2017communication} (FL, Figure~\ref{fig:cdsvsfl_graph}B)  a framework that allows deep learning models to be distributed and trained on local data, aggregating only their parameters in a central server.
The central server only ever sees a complex representation of the initial data, as learned by the local models, and those representations are combined by an algorithm called \textit{FederatedAveraging} before being redistributed for subsequent training.

Privacy preserving federated learning systems for medical image analysis have been mainly explored in the context of segmentation for brain~\cite{bakas2018identifyingBRATS, federatedbrainfmri2020, federatedbraintumor2019, roy2019braintorrent}, prostate \cite{sarma2021federatedPROSTATE} and COVID-19 affected regions \cite{COVIDkumar2020blockchain, yang2021federatedCOVID, liu2020experimentsCOVID}. 
However, diagnosis is a more challenging topic, and segmentation steps are often a prerequisite for diagnosis models to be accurate. The data scarcity problem is even more prominent in this case, as the ground truth information for the presence of disease is of a much more sensitive nature, and clinical registries are lacking. Label imbalance and demographic variabilities become more relevant as well, as different institutions usually contain different types of diseases, while in segmentation the expected ground truth is the same in all cases---i.e. a mask of the segmented parts. 
Among other things, segmentation allows for more flexible methods of data augmentation (even GAN-based generation) while in diagnosis augmentation should be done with extreme care so as to not shift the true value of the corresponding label (often requiring human-expert validation depending on the sophistication of the data generation / augmentation method, lest it risks adding noise). Segmentation also has the advantage of freely using 2D slices and patch-based approaches, which allows one to extract many training samples from a single scan, while in diagnosis, a single 3D volume (or sometimes series of longitudinal data) is used as a single training data point. 

For these reasons, related research in domains other than segmentation has been more limited, with studies tangential to diagnosis popping up only as early as 2020 in breast density classification \cite{BREASTroth2020federated}, and lung tumor survival prediction \cite{LUNGzerka2020blockchain}. With the emergence of the COVID-19 pandemic, the urgent need for diagnostic tools circumvented common obstacles and allowed researchers to collaborate in federated learning diagnosis for the first time~\cite{qayyum2021collaborativeCOVIDdiagnosis, zhang2021dynamicCOVIDdiagnosis}. Very recently, an open-source framework that integrates FL along with the functionality of end-to-end encryption to protect against inversion attacks was developed, trained and tested on paediatric X-ray classification~\cite{kaissis2021end}.

In this work, we focus on the diagnosis of cardiovascular disease (CVD) based on Cardiac MRI. The importance of furthering our understanding on the structure and function of the heart is highlighted by the prevalence of CVD in the population, which accounts for one third of annual deaths~\cite{ritchie2018causes, wilkins2017european}. Cardiac MRI has been the go-to modality for this task, allowing the assessment and delineation of the three heart segments---i.e. the myocardium, and the left and right ventricles---to identify the presence of anomalies such as myocardic infarctions or cardiomyopathies.
Based on this modality, and by leveraging hand crafted features, Machine Learning (ML) diagnostic tools have been developed with some success in single-center datasets~\cite{leiner2019machine, martin2020image}.

In terms of the Cardiac MRI modality, there has been a lot of literature on deep learning-based segmentation~\cite{zhang2019deep, luo2016novel, isensee2017automatic, jang2017automatic, zotti2017gridnet, patravali20172d, baumgartner2017exploration, rohe2017automatic, yang2017class} with diagnosis typically being a follow-up step, leveraging the segmentation masks and utilizing models such as random forests~\cite{khened2017densely, wolterink2017automatic}, support vector machines~\cite{cetin2017radiomics} or a simple diagnostic rule~\cite{liu2020residual}. These diagnosis models focus on two timepoints of the cardiac MRI per-patient: the phase of End-Diastole (ED) (maximum heart relaxation) and the phase of End-Systole (ES) (maximum heart contraction). Current such studies have emphasized on the the ACDC dataset~\cite{ACDCbernard2018deep}, a single center dataset hosting 100 subjects and five labels (20 subjects each). 

Khened et al.~\cite{khened2017densely} reported promising results on multi-label diagnosis but evaluated on a limited hold-out test set of 10 samples.
On the same task, Cetin et al.~\cite{cetin2017radiomics} used an SVM on top of radiomic features derived from manual segmentations, while 
Wolterink et al.~\cite{wolterink2017automatic} evaluated a random forest classifier instead, both using a cross validation scheme.
Liu et al.~\cite{liu2020residual} used an automated deep-learning based segmentation scheme, following up with a diagnostic rule.
Despite the impressive classification performance reported in these studies, these models are both trained and evaluated solely on a single-center dataset (ACDC), and thus, no assumptions can be made regarding their generalization to unseen centers and larger datasets. To deploy such models in the real world, one has to assume that new subjects being tested are IID to those seen during training and evaluation. In the domain of medical imaging, where data is highly heterogeneous between centers, this assumption is far from true~\cite{kwong2013need}. 
Furthermore, as these studies focus on a single center, no privacy preservation measures are studied, which are otherwise necessary for deployment of such models.

Despite the widespread interest in automated CMR diagnosis methods, multi-centric and distributed learning studies in the field are currently lacking. In this paper, we conduct our study with four centers, three of which were derived from the M\&M dataset \cite{victor_m_campello_2020_3886268}, and the fourth being a subset of  ACDC\cite{ACDCbernard2018deep}. We test the CDS and FL collaborative learning frameworks. As FL trains local models in each center, multi-label classification is a very challenging problem in the case of the M\&M dataset where many labels have little to no overlap between the centers, thus causing local models to overfit on different tasks. For this reason, we focus only on diagnosing hypertrophic cardiomyopathy (HCM)---i.e. a binary classification between normal (NOR) subjects and subjects suffering from HCM. HCM is the most common heritable cardiomyopathy, occurring in approximately as 0.29\%--–i.e. 1:344---of the adult population~\cite{marian2017hypertrophic, geske2018hypertrophic}.
Contrary to previous work on cardiac MRI diagnosis, our main goal here is to test the feasibility of cardiac MRI diagnosis in between multiple centers and highlight the importance of evaluating both IID and non-IID performance (i.e. testing models on partitions of the centers seen during the training and also on unseen centers) in a principled manner. 

Previous work in federated learning diagnosis on COVID-19~\cite{zhang2021dynamicCOVIDdiagnosis, qayyum2021collaborativeCOVIDdiagnosis} and paediatric X-ray classification~\cite{kaissis2021end} has focused on the development of state of the art federated learning frameworks---the latter one open-sourcing their pipeline which also integrates an encryption mechanism. In this study, we focus on the effects multi-center data has on this frameworks, conducting a systematic comparative analysis between the CDS and FL paradigms, testing a variety of data curation and augmentation techniques. By evaluating in two distinct cross-validation set-ups and repeating the experiments multiple times, we gain a robust estimate of both IID and non-IID performance, showcasing the gap between the two.
Our model follows the notoriously hard to train 3D-CNN architecture\cite{overfitting3DCNNs}, leveraging transfer learning by using an instance of the network that has been pretrained on action recognition.
Concretely, our contributions can be summarized as follows:
\begin{itemize}
    \item We present, to the best of our knowledge, the first federated learning study on CMR diagnosis and demonstrate that FL performance is comparable to CDS, while preserving patient privacy. 
    \item We propose a technique of inducing different priors to the model by leveraging the ground truth masks, illustrating an effective way to constrain the solution space and improve performance for deep learning-based multi-center CMR diagnosis in both collaborative learning set-ups.
    \item We apply a diverse set of data augmentations to artificially increase the data size and study their effect in a principled way on the collaborative learning frameworks, repeating the experiments with different CNN weight initializations to gain an estimate of model robustness~\cite{glorot2010understanding}.
    We also test a variation of the FL algorithm in this context, which assigns an equal vote to all centers in the training data and show that it is beneficial in some cases.
    \item Finally, by using two distinct repeated cross validation set-ups---one that uses a part of all centers as test set per fold and another that uses a whole center as test set per fold---we get an estimate of both on-site and out-of-site performance, showing that the two differ substantially and highlighting their importance for future diagnosis studies.
    \item To boost future research in the field, we make our code available for the research community. 
\end{itemize}

\section{Methods}

\subsection{Dataset}
We base our experiments on a unique dataset derived from a combination of M\&M (a multi-centric dataset gathered in a coordinated effort under the EuCanShare project~\cite{victor_m_campello_2020_3886268}) and ACDC (a single center dataset presented as a challenge in 2018~\cite{ACDCbernard2018deep}). Both datasets are composed of T1-weighted Cardiac Cine MRI sequences. M\&M consists of 6 centers and 4 labels corresponding to dilated cardiomyopathy (DCM), hypertrophic cardiomyopathy (HCM), abnormal  right  ventricle  (RV)  and  subjects  without  cardiac disease (NOR)\cite{victor_m_campello_2020_3886268}. 
Because of severe label imbalance between centers training a federated model on multi-label classification becomes a complex problem as local models overfit to a subset of the labels or even a single label (for example one center only has DCM cases). 
In our experiments, we consider the task of binary classification (HCM vs NOR), using a subset of M\&M in which the chosen labels are most balanced. Thus, we use 3 centers from the M\&M dataset, namely Sagrada Familia, Vall d'Hebron and SantPau and complement them with a subset of the ACDC dataset as a $4^{th}$ center. The final form of the dataset we used is outlined in Table~\ref{tab:dataset}.



\begin{table}[tb]
\caption{Dataset description, including meta-data and the class distribution of Normal (NOR) and hyperthropic cardiomyopathy (HCM) for the selected subset from the M\&M and ACDC datasets used in this study.}\label{tab:dataset}
\centering
\begin{tabular*}{\textwidth}{l@{\extracolsep{\fill}}ccc|ccc}
\toprule
Center          & Vendor & \small{Spatial Resolution (mm)} & \small{Slice Thickness (mm$^2$)}& NOR & HCM & Total  \\
\midrule
Vall d'Hebron   & Philips & 1.1516-1.2362 & 10.0 & 21  & 25  & 46 \\
Sagrada Familia & Siemens & 0.9765-1.6200 & 8.0-10.0 & 33  & 37  & 70   \\
SantPau         & Canon & 0.7955-1.8228 & 10.0 & 14  & 10  & 24   \\
ACDC            & Siemens & 1.3400-1.6800 & 5.0-10.0 & 20  & 20  & 40   \\
\midrule
Total           & & & & 88  & 92  & 180\\
\bottomrule
\end{tabular*}
\end{table}

\subsection{Data Preprocessing}
\label{sec:preprocessing}

\begin{figure}[tb]
\begin{center}
\includegraphics[width=\textwidth]{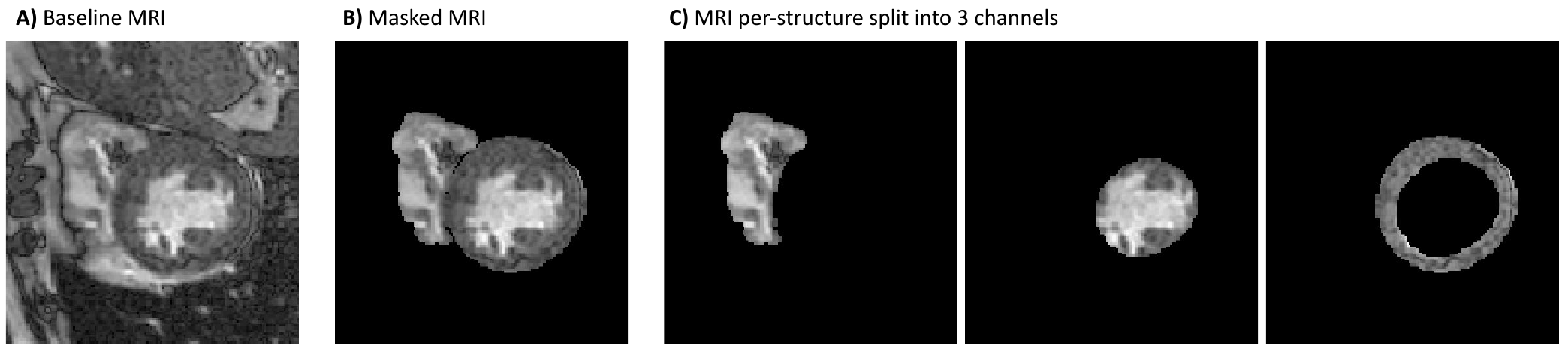}

\caption{An example of the induced priors used in this study: A) a baseline that is the 1-channel cardiac MRI, B) the baseline multiplied by the segmentation mask, C) the baseline split into three channels, one for each part of the heart (right ventricle, myocardium, left ventricle).
}
\label{fig:priors}
\end{center}
\end{figure}

As a first step, we apply N4 bias field correction to remove non-uniformity of low frequencies inherent to MRI~\cite{tustison2010n4itk}. Then we resample the volumes to a common spacing of 1$\times$1$\times$1 mm$^3$ on the entirety of the data used. 
The original spacing of the images vary~\ref{tab:dataset}, but all centers trim their original spacing, with the only exception of SantPau that in many cases interpolates its spatial resolution to reach a  1$\times$1$\times$1 mm$^3$ spacing. We crop the volumes using a 150$\times$150$\times$10 voxel window, centered on the center of the bounding box of non-zero values on their corresponding segmentation masks.

\subsubsection{Induced priors} 
\label{sec:priors}
Leveraging the segmentation masks that are provided by clinicians as part of the M\&M and ACDC datasets, we consider the following set-ups for our dataset: 1) a baseline that is the 1-channel MRI; 2) the MRI image multiplied by the segmentation mask (thus inducing a shape prior); and 3) the MRI image split into three channels, corresponding to left ventricle, right ventricle, and myocardium, inducing a strong prior of the heart structure (see Figure~\ref{fig:priors}). Notably, the initial layer of ResNet3D, was pretrained on action recognition RGB images and thus expects a 3-channel input. For this reason, the baseline and masked MRIs are copied three times before being fed to the model. Reinitializing the initial layer to a 1-channel input results in substantially worse performance, which, for brevity, we don't report in this paper.
Two timepoints are extracted from the 3D volumes corresponding to the ED and ES phases. The two timepoints are fed as separate samples to the network so as to leverage more data; however, the two timepoints are always in the same set (train, validation, test).

\subsubsection{Data Harmonization}  
\label{sec:harmonization}
A main obstacle when learning from multi-center MRI data is the inter-site variability, which is present even when the same acquisition protocol is used~\cite{nyholm2013variability, mirzaalian2015harmonizing}. To adapt the image intensities on the same scale, we apply Ny\'ul histogram matching~\cite{nyul2000new} followed by a rescaling to [0, 1].
Concretely, we match the histogram of each image to the average histogram of all images from the training centers.
As averaging is an aggregate query to the local data, it does not require access to individual subject data and thus preserves privacy. The final average histogram $H_{average}$ used for the histogram matching process is calculated as: 

$$H_{average}= \sum_{k=1}^K \frac{N_k}{N}\sum_{n=1}^{N_k} H_n^k,$$
where $K$ is the total number of centers, $N_k$ the number of samples for center $k$, and $N$ the total sample size of all centers. 

In the majority of our experiments, we crop our inputs with segmentation masks, and thus only the values within the mask are used for the histogram matching as well. As we use two distinct evaluation set-ups, in some cases not all centers are calculated in the average. This is further clarified in section \ref{sec:evaluation} 
where these set-ups are also explained.

\begin{figure}[tb]
\begin{center}
\includegraphics[width=\textwidth]{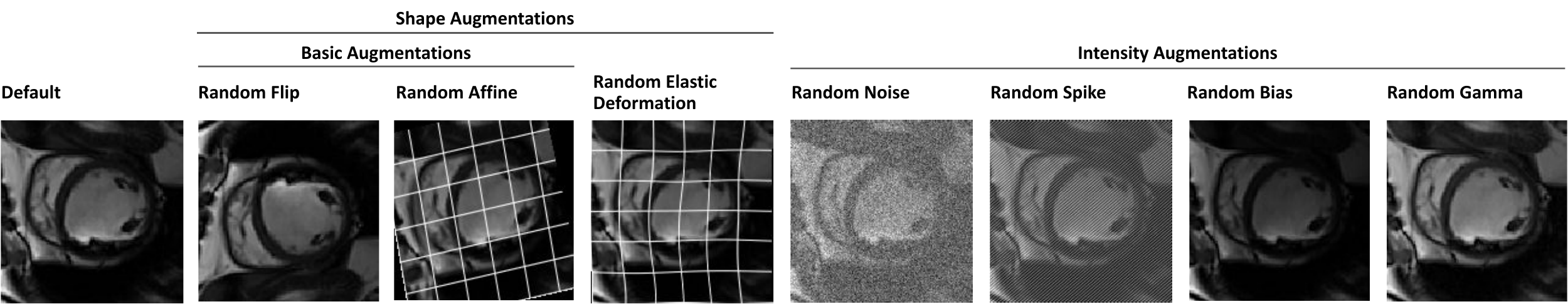}

\caption{Illustration of augmentation techniques used in our analysis on a single CMR image slice.} 
\label{fig:augmentations}
\end{center}
\end{figure}
\subsection{Data Augmentation}
\label{sec:dataAugmentation}
Due to the limited number of samples in the M\&M dataset, we test several data augmentation techniques to artificially increase the size of the training set. These techniques also apply domain shift effects and make the representations learned more invariant to certain features, which can improve generalization to unseen datasets~\cite{castro2020causality}.  To test how this dataset shift affects our framework, we compare four different augmentation set-ups and repeat the experiments both for FL and CDS. These set-ups are as follows:
\begin{itemize}
    \item No Augmentations.
    \item Basic Augmentations: affine transformations (rotation by varying degrees), horizontal and vertical flipping in axial view.
    \item Shape Augmentations: includes the basic augmentations plus elastic deformation.
    \item Shape and Intensity Augmentations: includes all aforementioned augmentations, plus random MRI spike artifacts, random MRI bias field artifact, noise sampled from a Gaussian with $\mu = 0$ and $\sigma \sim (0,0.25)$, random gamma transformations (randomly changes contrast of an image by raising its values to a random power within a specified range).
\end{itemize}
During training on one of these set-ups (except the No Augmentations set-up), an augmentation is sampled from the aforementioned pools and is applied on the input data with a probability of 50\%. That means that every image has a 50\% chance of being augmented.
These types of data augmentations are known to be effective in improving a model's generalization as they introduce domain shift effects. However, the chosen augmentations are also reflective of clinical reality as we avoid extreme cases of elastic deformation and consider the random noise augmentation as representative of bad acquisitions, which often escape quality control.
All augmentations used and studied in this work were obtained using the TorchIO library \cite{perez-garcia_torchio_2020} and are illustrated in Figure~\ref{fig:augmentations}.

\subsection{3D-CNN Model}
\label{sec:3dcnn}

3D-CNNs have the potential to retrace the success story of 2D-CNNs; however, their immense size is cause of two major drawbacks---i.e. a high computational cost and the curse of dimensionality which causes them to overfit \cite{overfitting3DCNNs}. In medical imaging, where data is very scarce, 3D-CNNs have been used successfully by applying transfer learning schemes, utilizing data from entirely different domains \cite{singh20203d}. In our case, we use the 3D-CNN ResNet18 model as defined by Tran et al.\cite{ResNet3DModel-tran2018closer} and, instead of initializing the weights randomly, we load an instance of the model that has been pretrained on the action recognition dataset Kinetics-400 \cite{DATASET-kay2017kinetics}. This is beneficial because the early layers of the network tend to extract similar features (such as edges or blobs) irrespective of the domain that are beneficial to all imaging tasks~\cite{kushibar2019supervised}.
To constrain the model from overfitting, we freeze the initial layers and train a newly initialized linear layer with a sigmoid activation function to the task of binary classification of HCM vs NOR. Concretely, out of the 33,166,785 parameters our network has in total, only the 512 parameters of the final linear layer are trained in this case. In our preliminary experiments, 256, 512, 1024 and 2048 channels on the linear layer were tested and 512 was found to perform best.

This, however, introduces a second problem: a strong bias on the model for a completely different domain and data modality. To find a proper trade-off between the issue of bias and overfitting, we experimented by fine tuning the pre-trained layers with varying learning rates (1e-4, 1e-5, 1e-6, 1e-7 and 0) while the newly initialized linear layer was always trained with a much higher learning rate of 0.01. Our preliminary experiments showed that for the pre-trained layers, a learning rate of 1e-5 performed best, and thus we used this configuration for all subsequent experiments.
The models are trained for a maximum of 100 epochs, stopping early if validation performance stagnates for 10 epochs. In our experiments, the early stopping occurred on the 20-30th epoch.

\subsection{Federated Learning}
\label{sec:FL}

We first initialize a global model and then distribute it across the four centers. It is fundamental that on every step, including the initialization the model being distributed is identical, otherwise the aggregation across models will result in a non-sense representation and training will not progress \cite{FEDERATEDORIGINmcmahan2017communication}. The models are trained for seven iterations---i.e. on seven batches of data---on each center and are aggregated after each epoch. To parse the entire data within the same number of iterations on each center, a different batch size is used.

After training locally, the models are aggregated using the FederatedAveraging algorithm \cite{FEDERATEDORIGINmcmahan2017communication}. Concretely, each model makes an update step in its respective center $k$ using a learning rate $\eta$ and the gradients $g_k$ so that $$w^k_{t+1} \longleftarrow w_{t} - \eta g_k, \forall  k $$ Then, these weights are aggregated to the global model in a way that is proportionate to the sample size of each center, $$W_{t+1} \longleftarrow \sum_{k=1}^K \frac{n_k}{n}w^k_{t+1},$$ where $n$ is the total sample size and $n_k$ the sample size of center $k$.

The reason that the FederatedAveraging algorithm weights the model parameters by a factor proportionate to the sample size of the center, is so that the better informed models outweigh the others \cite{FEDERATEDORIGINmcmahan2017communication}. However, medical centers are liable to domain shift effects due to different scanners, acquisition protocols, and different demographics from center to center. Given that data from different centers tends to be severely imbalanced (in our case Sagrada Familia has three times the size of SantPau), weighting the models by the sample is liable to add severe bias towards some cases over others. Thus we also tested a modified version of the FederatedAveraging algorithm where each model gets an equal weight during aggregation, which we will be refering to as FL-EV (as in Equal Voting). In this case the averaging equation becomes: $$W_{t+1} \longleftarrow \frac{1}{K}\sum_{k=1}^K w^k_{t+1}.$$

\subsection{Evaluation Procedure}
\label{sec:evaluation}
\begin{figure}[t]
\centering
\includegraphics[width=0.95\linewidth]{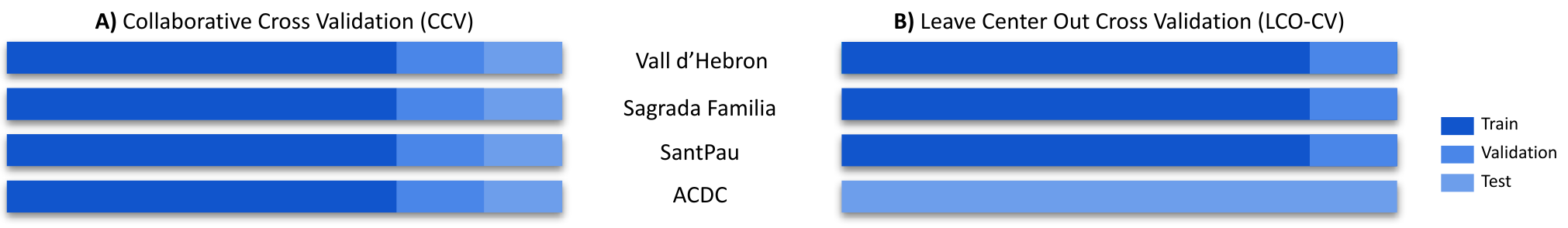}
\caption{An illustration of a single iteration for: A) Collaborative Cross Validation (CCV), where centers coordinate their splitting across 5 folds so that each center provides 20\% of its data as test set and the rest as training and validation, B) Leave Center Out Cross Validation (LCO-CV) which runs for as many iterations as there are centers in the dataset, each time using a different center as test set, and the rest as training and validation.}
\label{fig:evaluation}
\end{figure}

As we are dealing with data of small size, simple hold-out methods are inefficient to accurately represent the performance of our models\cite{hastie2009elementsCROSSVALIDATION}. Thus, all models are validated under a 5-fold cross validation scheme that utilizes the entire dataset. Concretely, we split the dataset into five folds, so that 20\% is unseen during the training procedure and is used as a test set. From the remaining 80\% we use 90\% for training and 10\% as validation. The purpose of the validation set is simply to find the early stop point. 

In the past, federated learning performance has been evaluated under a collaborative cross validation (CCV) set-up~\cite{bakas2018identifyingBRATS}. In this set-up, cross validation is coordinated across centers so that each center splits its data into train, validation and test sets using the same percentage allocation. As a result, the final test set of each fold is an aggregate of different centers. 
However, when dealing with multi-center data, one has to take into account that a model should be deployable to centers unseen during the training and an estimate of out-of-site performance is required. For this reason, we are also estimating how our model generalizes to unseen centers under a different set-up that is Leave Center Out Cross Validation (LCO-CV). In this case, the data is split into as many folds as there are centers present (in our case 4 folds) and each fold uses a different center as testing set. This cross validation scheme has been used before in other distributed learning studies \cite{deist2017infrastructure}. In both CCV and LCO-CV cases, FL and CDS are using the exact same split of the data so that all folds are comparable. A schematic of this evaluation procedure is presented in Figure~\ref{fig:evaluation}. 
For the CDS set-up, at each epoch a single model is trained and then evaluated. For the FL set-up, however, the evaluation process is more complex: 1) train the local copies of the model for one epoch; 2) send the local models to the server; 3) aggregate all local models to obtain an updated global model; 4) send copies of the updated global model to all centres; 5) evaluate the model on each centre locally. In this study, steps 2 and 4 are simulated within a single server; in the real world, however, these are the points where communication bottlenecks would occur due to possible bandwidth limitations~\cite{FEDERATEDORIGINmcmahan2017communication}. 
The best model based on validation performance is used to predict the test set. After the entire cross validation procedure is complete, the Area Under the Receiver Operating Characteristic Curve score (AUC) is calculated based on the predictions derived from each respective test set. 

In the case of CCV, to ensure fairness in comparison, the same folds are used for the FL and the CDS scheme. We repeat each experiment five times---i.e. using five different network weight initalizations. 
The reason for this is that a DL model's performance is liable to vary, as different initializations tend to converge to different optima~\cite{glorot2010understanding}. By repeating each experiment five times and calculating the average AUC across repetitions, we obtain a more accurate estimate of the true performance and an estimate of the framework's robustness.

\begin{figure}[t]
\centering
\includegraphics[width=0.33\linewidth]{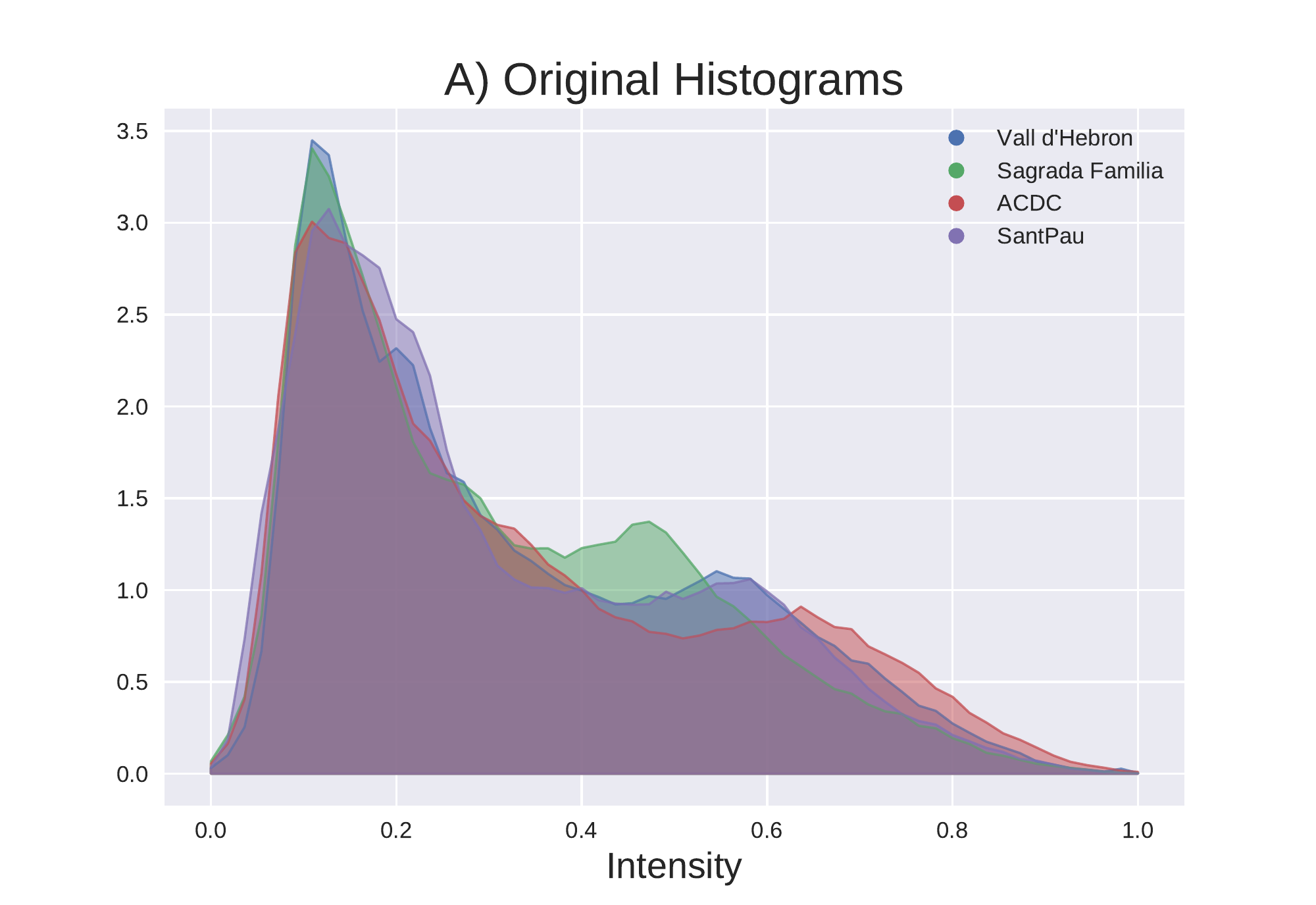}
\includegraphics[width=0.33\linewidth]{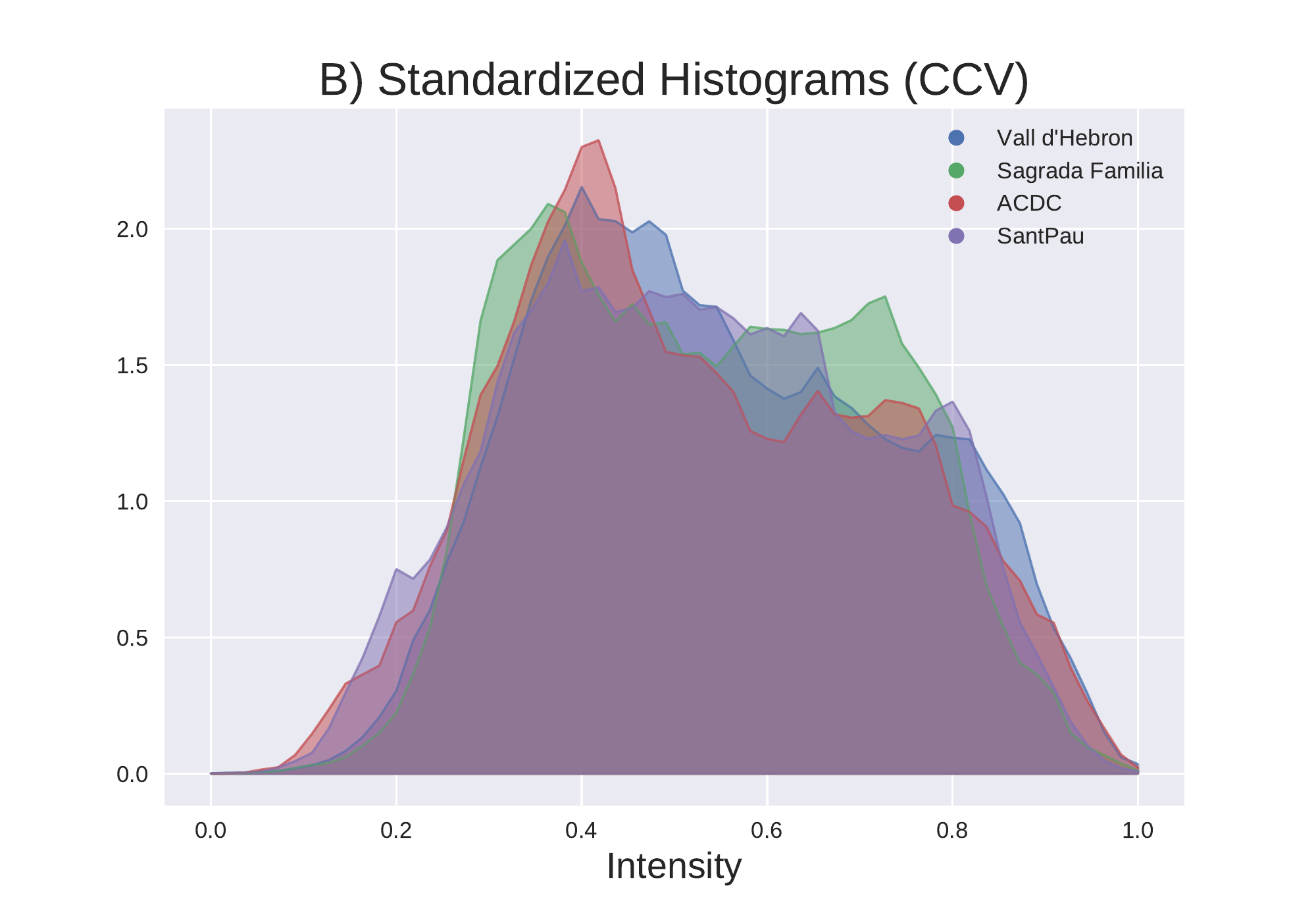}
\includegraphics[width=0.33\linewidth]{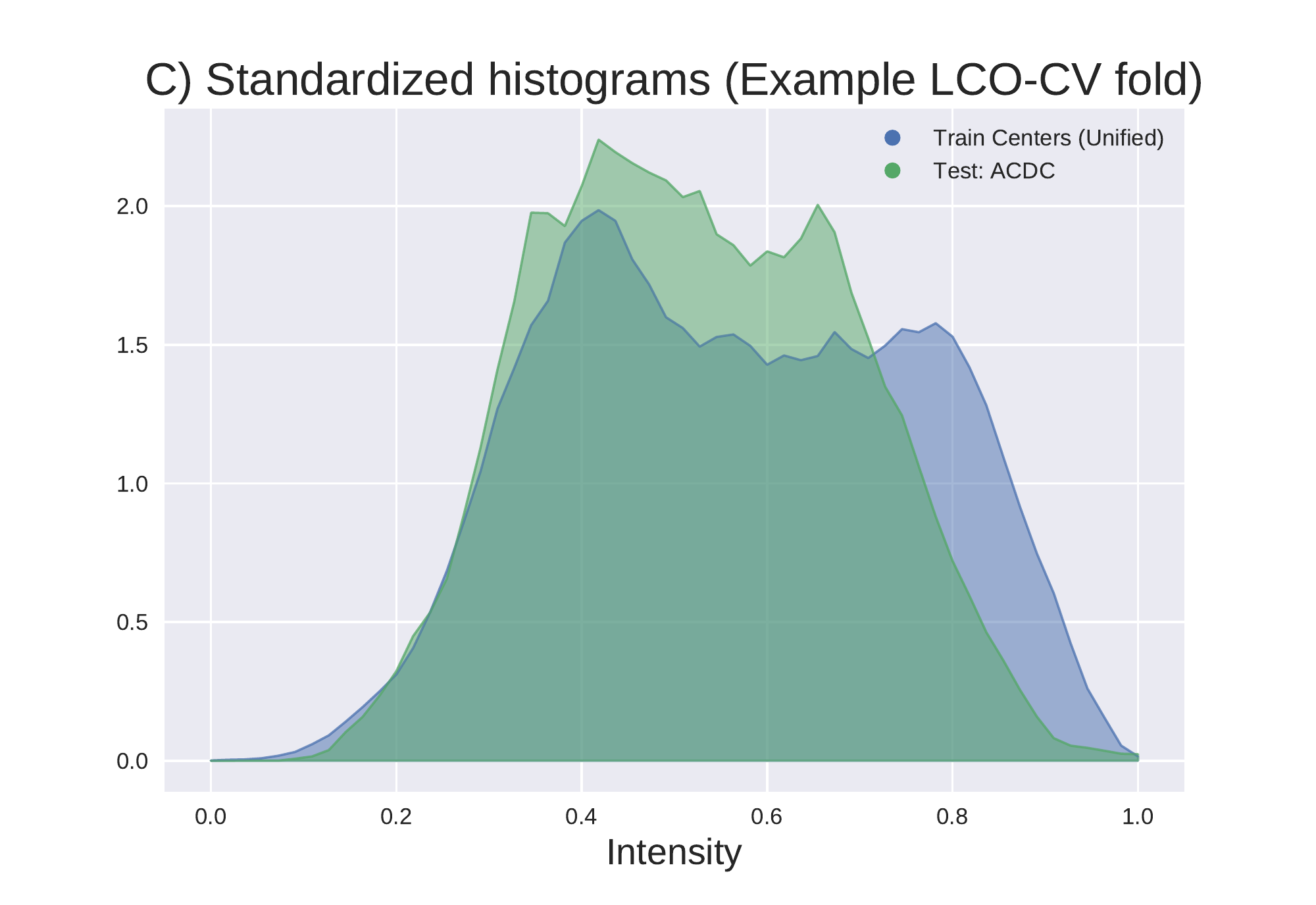}
\caption{Histograms of different center intensities before and after histogram standardization. The two evaluation techniques imply different adaptation set-ups as for the \textit{CCV} we are standardizing using an average from all of the centers, while for \textit{LCO-CV} we consider one center entirely unseen.}
\label{fig:histogramsCCV}
\end{figure}

Regarding data harmonization, in the LCO-CV, the reference histogram used from matching does not include the test center, thus making the task harder when compared to CCV where all centers are used for the derived average. The histograms of each center before and after the standardization are outlined in Figure~\ref{fig:histogramsCCV}.
In this diagram, one example of an \textit{LCO-CV} iteration is visualized where Vall d'Hebron is the test set. Vall d'Hebron in this case is matched to the histogram average as calculated by the train centers. Notably, the test set becomes more challenging because of this limitation, and in this example we can see this by the presence of a second peak on Vall d'Hebron's intensity distribution.

\begin{figure}[t]
\begin{center}
\includegraphics[width=\textwidth]{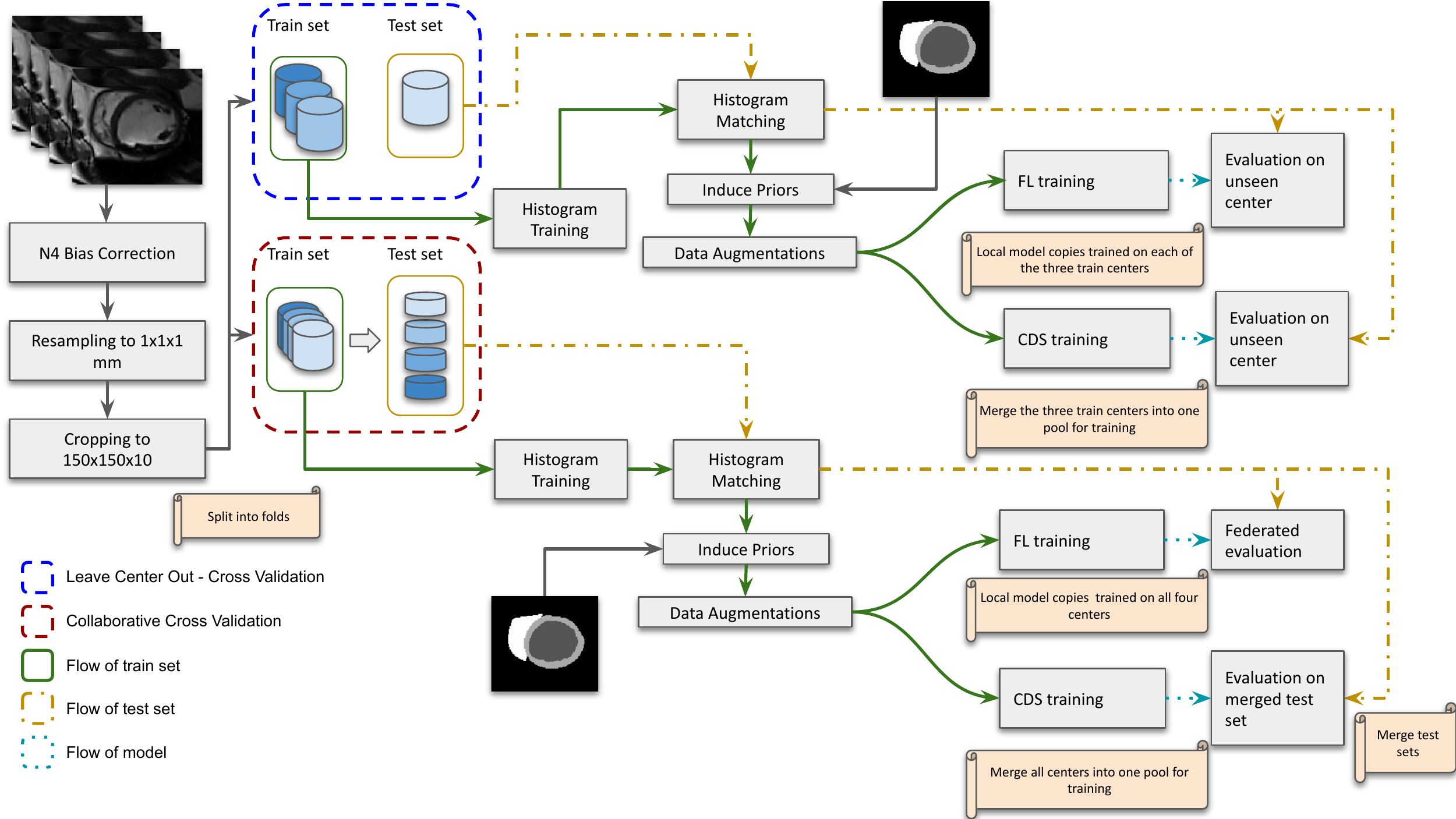}

\caption{An overview of the pipeline used for the experiments in this study.}
\label{fig:pipeline}
\end{center}
\end{figure}

The overall pipeline is summarized in Figure~\ref{fig:pipeline}.

\section{Results}

In the initial experiments, we compare the baseline data to a curated version with induced shape priors as visualized in Figure~\ref{fig:priors}. The results are outlined in Table~\ref{tab:curation}. For both collaborative learning frameworks, performance improves as we induce a shape prior (Masked MRI) and peaks when we use the per-structure split---a prior of the heart's structure.
FL exhibits similar performance to CDS in all cases, however the standard deviations are much lower in the case of FL. Moreover, the difference in performances of CDS and FL decreases after constraining the solution space using the shape priors.

\begin{table}[tb]
\caption{AUC results for different curation types for Collaborative Data Sharing (CDS) and Federated Learning (FL). The reported numbers are calculated across five repeated experiments with different seeds.}
\centering
\begin{tabular*}{\textwidth}{l@{\extracolsep{\fill}}cccc}
\toprule
\textbf{Curation Type} & Baseline MRI & Masked MRI & MRI per-structure split    \\
\midrule
\textbf{CDS}        & 0.727$\pm$0.0298        & 0.810$\pm$0.0120  & \textbf{0.856$\pm$0.011}      \\
\textbf{FL}        & 0.747$\pm$0.0005           & 0.826$\pm$0.0005   & \textbf{0.861$\pm$0.003}     \\
\bottomrule
\end{tabular*}
\label{tab:curation}
\end{table}

\begin{table}[tb]
\caption{Collaborative Cross Validation (CCV) performance evaluated using the AUC metric. Each number represents the average AUC, calculated across five repeated experiments with different seeds.}
\centering
\begin{tabular*}{\textwidth}{l@{\extracolsep{\fill}}ccccccc}
\toprule
\textbf{Augmentations}                     & \textbf{Framework} & \textbf{Vall d'Hebron} & \textbf{Sagrada Familia} & \textbf{ACDC}          & \textbf{SantPau}       & \textbf{Total}         \\
\midrule
\multirow{3}{*}{\textbf{none}}             & \textbf{CDS}       & 0.898$\pm$0.004          & 0.862$\pm$0.008            & \textbf{0.797$\pm$0.026} & 0.853$\pm$0.005          & 0.856$\pm$0.011          \\
                                           & \textbf{FL}        & \textbf{0.942$\pm$0.002} & 0.896$\pm$0.001            & 0.799$\pm$0.003          & \textbf{0.867$\pm$0.005} & \textbf{0.861$\pm$0.003} \\
                                           & \textbf{FL-EV}     & \textbf{0.941$\pm$0.002} & 0.874$\pm$0.003            & 0.803$\pm$0.002          & \textbf{0.875$\pm$0.002} & \textbf{0.852$\pm$0.002} \\
\midrule
\multirow{3}{*}{\textbf{basic}}            & \textbf{CDS}       & \textbf{0.861$\pm$0.014} & 0.784$\pm$0.012            & \textbf{0.778$\pm$0.041} & 0.845$\pm$0.016          & 0.809$\pm$0.021          \\
                                           & \textbf{FL}        & \textbf{0.927$\pm$0.002} & \textbf{0.872$\pm$0.002}   & 0.798$\pm$0.001          & \textbf{0.807$\pm$0.001} & \textbf{0.844$\pm$0.002} \\
                                           & \textbf{FL-EV}     & 0.933$\pm$0.001          & \textbf{0.859$\pm$0.001}   & 0.774$\pm$0.001          & \textbf{0.818$\pm$0.004} & 0.835$\pm$0.002          \\
\midrule
\multirow{3}{*}{\textbf{shape}}            & \textbf{CDS}       & 0.897$\pm$0.006          & 0.810$\pm$0.023             & 0.770$\pm$0.024           & 0.803$\pm$0.017          & \textbf{0.827$\pm$0.018} \\
                                           & \textbf{FL}        & 0.933$\pm$0.002          & \textbf{0.871$\pm$0.004}   & \textbf{0.826$\pm$0.001} & \textbf{0.901$\pm$0.002} & \textbf{0.848$\pm$0.002} \\
                                           & \textbf{FL-EV}     & 0.916$\pm$0.001          & 0.852$\pm$0.001            & 0.825$\pm$0.001          & 0.883$\pm$0.001          & 0.839$\pm$0.001          \\
\midrule
\multirow{3}{*}{\textbf{shape and intensity}} & \textbf{CDS}       & \textbf{0.897$\pm$0.009} & 0.821$\pm$0.008            & \textbf{0.859$\pm$0.028} & 0.833$\pm$0.025          & \textbf{0.849$\pm$0.018} \\
                                           & \textbf{FL}        & 0.905$\pm$0.001          & 0.886$\pm$0.001            & 0.785$\pm$0.005          & 0.858$\pm$0.003          & 0.839$\pm$0.003          \\
                                           & \textbf{FL-EV}     & 0.917$\pm$0.001          & \textbf{0.870$\pm$0.002}    & 0.800$\pm$0.003            & \textbf{0.880$\pm$0.002}  & 0.842$\pm$0.002\\
\bottomrule
\end{tabular*}
\label{tab:CCV}
\end{table}

\begin{table}[tb]
\caption{Leave Center Out (LCO-CV) performance evaluated using the AUC metric. Each number represents the average AUC calculated across five repeated experiments with different network initializations.}
\centering
\begin{tabular*}{\textwidth}{l@{\extracolsep{\fill}}ccccccc}
\toprule
\multicolumn{1}{l}{\textbf{Augmentations}} & \textbf{Framework} & \textbf{Vall d'Hebron} & \textbf{Sagrada Familia} & \textbf{ACDC}          & \textbf{SantPau}       & \textbf{Total}         \\
\midrule
\multirow{3}{*}{\textbf{none}}             & \textbf{CDS}       & 0.870$\pm$0.020          & 0.809$\pm$0.010             & \textbf{0.616$\pm$0.064} & 0.784$\pm$0.026          & 0.732$\pm$0.008          \\
                                           & \textbf{FL}        & \textbf{0.897$\pm$0.002} & 0.815$\pm$0.001              & 0.599$\pm$0.002          & \textbf{0.835$\pm$0.001} & 0.746$\pm$0.001          \\
                                           & \textbf{FL-EV}     & \textbf{0.894$\pm$0.001} & 0.816$\pm$0.001            & 0.566$\pm$0.002          & \textbf{0.837$\pm$0.001} & \textbf{0.779$\pm$0.004} \\
\midrule
\multirow{3}{*}{\textbf{basic}}            & \textbf{CDS}       & \textbf{0.916$\pm$0.014} & 0.816$\pm$0.016            & \textbf{0.654$\pm$0.068} & 0.776$\pm$0.019          & 0.759$\pm$0.016          \\
                                           & \textbf{FL}        & \textbf{0.922$\pm$0.002} & \textbf{0.855$\pm$0.001}   & 0.554$\pm$0.002          & \textbf{0.818$\pm$0.001} & \textbf{0.791$\pm$0.001} \\
                                           & \textbf{FL-EV}     & 0.886$\pm$0.007          & \textbf{0.868$\pm$0.001}   & 0.532$\pm$0.003          & \textbf{0.810$\pm$0.001}  & 0.773$\pm$0.005          \\
\midrule
\multirow{3}{*}{\textbf{shape}}            & \textbf{CDS}       & 0.900$\pm$0.024          & 0.834$\pm$0.028            & 0.641$\pm$0.084          & 0.796$\pm$0.013          & \textbf{0.764$\pm$0.022} \\
                                           & \textbf{FL}        & 0.861$\pm$0.002          & \textbf{0.879$\pm$0.001}   & \textbf{0.668$\pm$0.003} & \textbf{0.845$\pm$0.007} & \textbf{0.766$\pm$0.001} \\
                                           & \textbf{FL-EV}     & 0.803$\pm$0.003          & 0.869$\pm$0.001            & 0.632$\pm$0.008          & 0.821$\pm$0.001          & 0.737$\pm$0.004          \\
\midrule
\multirow{3}{*}{\textbf{shape and intensity}} & \textbf{CDS}       & \textbf{0.901$\pm$0.031} & 0.829$\pm$0.013            & \textbf{0.743$\pm$0.048} & 0.793$\pm$0.018          & \textbf{0.776$\pm$0.008} \\
                                           & \textbf{FL}        & 0.887$\pm$0.001          & 0.807$\pm$0.003            & 0.472$\pm$0.003          & 0.840$\pm$0.003           & 0.731$\pm$0.009          \\
                                           & \textbf{FL-EV}     & 0.892$\pm$0.001          & \textbf{0.854$\pm$0.005}   & 0.471$\pm$0.002          & \textbf{0.844$\pm$0.001} & 0.768$\pm$0.003         \\
\bottomrule
\end{tabular*}
\label{tab:LCOCV}
\end{table}

\begin{figure}[t]
\centering
\includegraphics[width=0.49\linewidth]{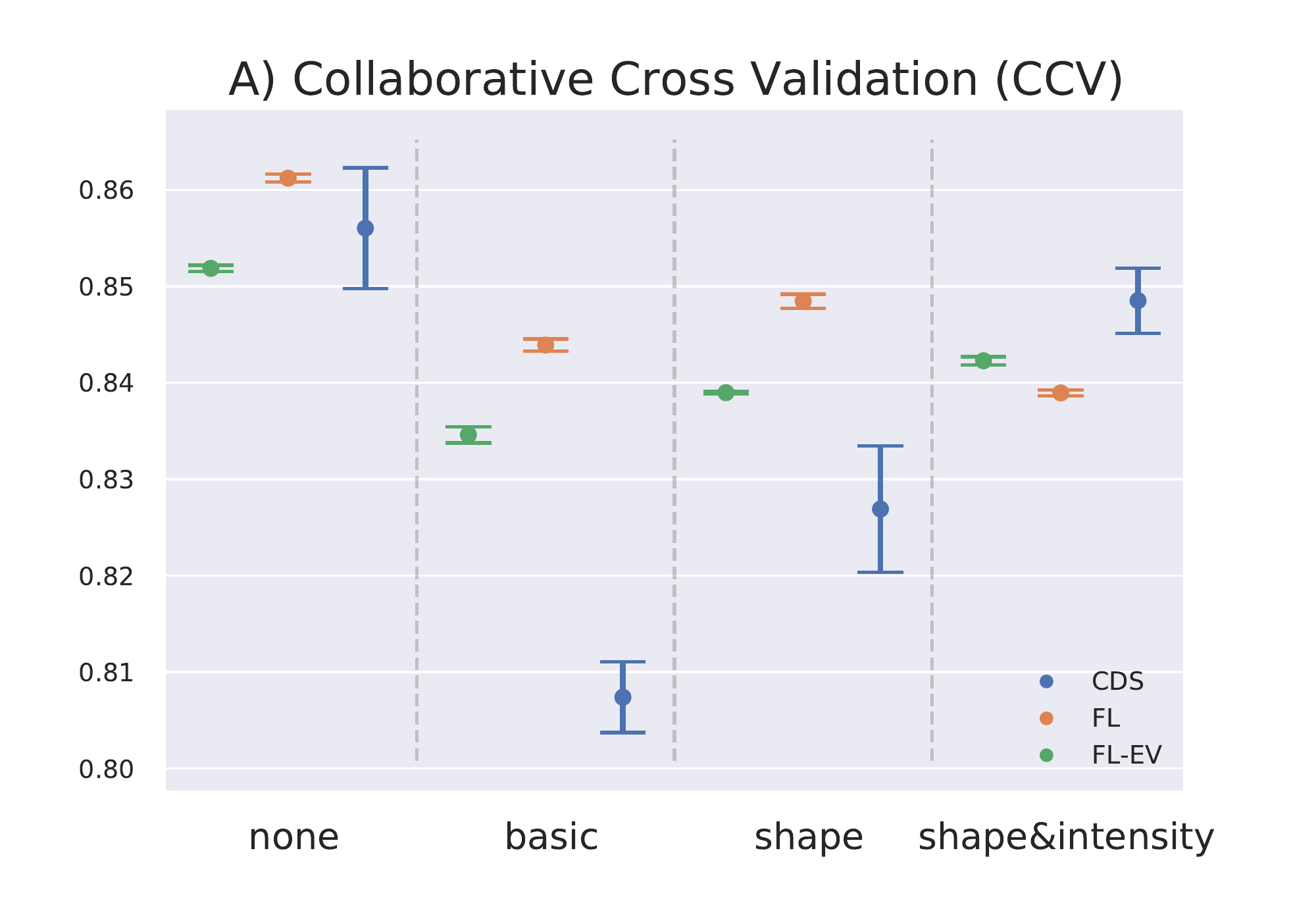}
\includegraphics[width=0.49\linewidth]{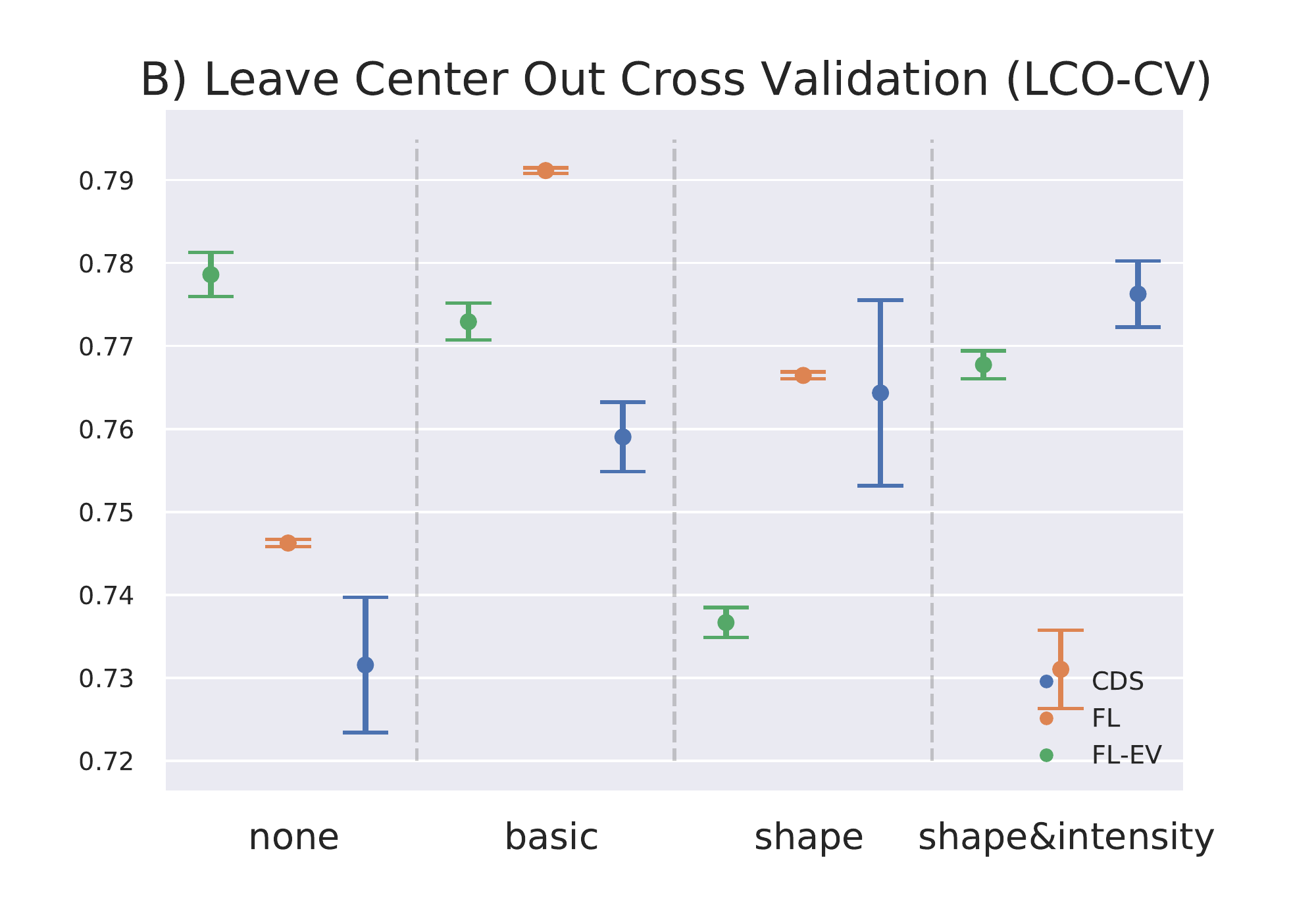}
\caption{CDS, FL and FL-EV are tested given four data augmentation set-ups and for each set-up the experiment is repeated with 5 different network initializations, both for the CCV and LCO-CV set-ups. AUC is evaluated across all folds and we obtain 5 AUC metrics per set-up whose error bars are displayed here.
}
\label{fig:AUCoverall}
\end{figure}

A systematic comparative study is conducted on the different augmentation set-ups using both the CCV and LCO-CV evaluation procedures. We find that CCV performance drops in all collaborative learning frameworks (Table~\ref{tab:CCV} and Figure~\ref{fig:AUCoverall}) once augmentations are introduced.
Since for CCV the training and testing data come from the same distribution, augmentations---which are only applied on the training partition---introduce domain shift between the two subsets. The results are indicative that this domain shift outweighs the benefit of the additional data.

In the case of LCO-CV, where the train and testing sets come from different distributions, CDS performance consistently improves from additional augmentations, reaching its highest when the most augmentations are applied (Figure~\ref{fig:AUCoverall}). FL also benefits from the basic augmentations of rotation and flipping, but its performance drops as shape and intensity augmentations are introduced, and CDS surpasses FL. Interestingly, models trained under an FL framework (either FL or FL-EV) exhibit incredibly consistent results across different initializations of the same set-ups, while CDS-framed models show a high amount of variance. This indicates that the federated approaches boost the robustness of these models.


To analyze the different centers individually, it is more interesting to focus on the LCO-CV set-up (Table~\ref{tab:LCOCV}). In this case, the out-of-domain performance is evaluated as the testing center in each fold is entirely unseen.
The most striking difference pertains to ACDC, where CDS outperforms the FL method in most cases, and by a margin of 0.3 on the AUC metric when intensity augmentations are introduced. 


In terms of the FL-EV scheme, Sagrada Familia---the largest of all centers in the used cohort---seems to be the only center to benefit from the scheme by a substantial margin. This occurs when the maximum amount of augmentations are introduced to the data, where FL-EV outperforms FL by a margin of 0.05 on Sagrada Familia, reflected in the total result.
 



\section{Discussion}

Automatic CMR diagnosis has thus far focused on isolated centers, with training happening locally and evaluation limited to IID subsets of the data~\cite{khened2017densely, wolterink2017automatic, liu2020residual, cetin2017radiomics}. We have presented the first federated CMR diagnosis study and showcased two distinct evaluation set-ups to quantify both IID and non-IID performance. The gap in performance between the two evaluation frameworks was found to be critical, with LCO-CV consistently displaying worse results by a margin of approximately 0.1 on the AUC metric. This clearly suggests that for such models to be reliable and eventually adopted clinically, the LCO-CV set-up needs to be adopted in future research to account for non-IID performance. The necessity of this is exemplified by the presence of bias against ACDC on the shape and intensity set-up where FL exhibits an AUC performance of about 0.85 to 0.89 for the M\&M centers but only 0.472 for ACDC (Table ~\ref{tab:LCOCV}). Although lower performance is also seen in the CCV case for ACDC, the magnitude is not as critical (0.785 on AUC, Table~\ref{tab:CCV}). As M\&M was part of a coordinated data collection within the EuCanShare project~\cite{victor_m_campello_2020_3886268} acquisition protocols were bound to be more similar than that of ACDC which was collected in 2018 as part of a challenge~\cite{ACDCbernard2018deep}. We believe this to be a possible explanation for the exhibited bias against this center.

Our results reveal a couple of interesting behaviors on federated learning in this use case.
Although FL has consistently been outperformed by CDS in medical imaging studies in the past, here FL outperforms CDS in the majority of our experiments. This result may come as surprising at first; however, FL has no inherent disadvantage over a CDS set-up. 
In the past, it has been shown that averaging the weights of copies of the same model in different time points 
boosts performance by approximating a lower point in the loss space \cite{izmailov2018averaging,garipov2018loss}. We believe that, although in the case of FL the averaging happens at the same timepoints, with copies trained on different datasets, a similar effect occurs and becomes apparent in the use case of CMR, possibly due to the low amount of data. Furthermore, averaging across models seems to result in a stabilizing effect, and a higher performance of FL across different initializations of the same model, (exhibited by our repeated experiments over the same set-ups, Figure \ref{fig:AUCoverall}).

Interestingly, as we artificially increase the size of the dataset with data augmentation techniques, CDS starts outperforming FL. Although FL first benefits from the basic augmentations of rotation and flipping, it exhibits a sharp decline once the more complex shape and intensity augmentations are realized. This stands as evidence that FL is more sensitive to the domain shift effects these augmentations have on the data in a way that overwhelms the benefit of having additional data.

On the CCV set-up, where all centers contribute to the test data, the performance is evaluated on the same domain as the training set. Thus, the augmentation techniques shift the distribution away from the train set in a manner that always hurts performance. LCO-CV evaluates performance on an unseen center and thus on an unseen domain. In this case, CDS performance is impacted positively increasing as more augmentations are applied. 

Additionally, we used an equal voting strategy as an alternative to the default sample-size based weighting of the FederatedAveraging algorithm. We labeled this variant FL-EV as it attributes Equal Voting to all centers, exhibiting deviations in performance---in some cases improving and in some cases worsening (Tables ~\ref{tab:CCV}, \ref{tab:LCOCV}). 
An interesting example of this is Sagrada Familia (Table~\ref{tab:LCOCV})  under shape and intensity augmentations, as it benefits from the equal voting by a margin of 0.05 on the AUC metric. When Sagrada Familia is the test set, SantPau's vote increases while Vall d'Hebron decreases (based on their sample sizes). Thus, the explanation for the observed differences could be that Sagrada Familia's data distribution is much closer to SantPau's than Vall d'Hebron's and benefits from the effect on votes. 
We believe this warrants further exploration in the future, studying a spectrum of intermediate voting schemes that are not entirely equal nor as imbalanced as the per-center sample size.

In conclusion, through extensive analysis and experiments, we demonstrated that, even with a small sample size of 180 subjects derived from four centers, federated learning for Cardiac MRI diagnosis achieves promising performance that is comparable to collaborative data sharing. We highlighted the importance of a principled evaluation that accounts for both in and out of site performance and showed how models trained under a federated learning framework exhibit increased robustness and can be more sensitive to domain shift effects. As different centers seem to benefit in different ways from the interplay of augmentations and the collaborative learning frameworks, we believe that further research is required to delineate the underlying factors. In the future, bigger datasets with a wider diversity of centers should be used to systematically analyze and further verify these effects. Furthermore, our study was constrained to binary classification, but in the future semi-supervised federated learning methods will be needed to integrate labels that don't fully overlap between centers to enable multi-label classification. Also, as we used segmentation masks supplied by clinicians, the current pipeline still relies on expert support to provide these segmentations and could be replaced by an automated federated segmentation pipeline, appended as a first step to the current pipeline.

There is still a lot of fields in automated image-based diagnosis where federated learning has yet to make a presence, including prediction of prevalent diseases like retinopathy, diabetes, neurological disorders and cases of cancer in breast, liver or colon. Tangential fields like that of survival and treatment outcome prediction that are still underdeveloped due to the lack of data would also stand to benefit from the impact of federated learning and warrant similar exploration.
We firmly believe that such studies will be a fundamental step to pave the way for multi-center studies going forward.

\bibliography{sample}



\section*{Acknowledgements}

This project has received funding from the European Union’s Horizon 2020 research and
innovation programme under grant agreement No 952103.

\section*{Author contributions statement}


A.L., K.K., P.G. and K.L. conceived the experiments,  A.L. conducted the experiments and created the software used in this study, 
A.L., K.K.,  and K.L. analyzed the results.  A.L., K.K., S.W. and K.L reviewed and edited the manuscript.

\section*{Additional information}

\textbf{Competing interests}
Sean Walsh declares the following financial interests/personal relationships which may be considered as potential competing interests: within and outside the submitted work, is the recipient of grants/sponsored research agreements in the areas of medical imaging, artificial intelligence, data science, applied to the clinical specialties of oncology and respiratory medicine. He holds a leadership position within Oncoradiomics SA, has shares in the company Oncoadiomics SA, and is co-inventor of submitted patents on behalf Oncoradiomics SA. The rest of the authors declare no competing interests.



\end{document}